\documentclass{aa}             % <-- "class" anstelle von "style"
\usepackage{amssymb,graphicx}  % <-- "graphicx" fuer Bilder, "amssymb"
\begin{document}

\thesaurus {06(02.01.2 - 02.02.1 - 13.25.5
 -  08.09.2 Cygnus X-1, Nova Muscae 1991 - 11.14.1)}

\title{Evaporation: The change from accretion via a thin disk to a
coronal flow}
\author{F. Meyer \inst{1}, B.F. Liu \inst{1,2}, E. Meyer-Hofmeister \inst{1}
}
\offprints{Emmi Meyer-Hofmeister\\ e-mail: emm@mpa-garching.mpg.de}
\institute{Max-Planck-Institut f\"ur Astrophysik, Karl
Schwarzschildstr.~1, D-85740 Garching, Germany
\and
Yunnan Observatory, Chinese Academy of Sciences. P.O.Box 110, Kunming 650011, China
} 

\date{Received:s / Accepted:}
\titlerunning {Evaporation from accretion disk to coronal flow}
\maketitle

\begin{abstract}

We present a model for a corona above a geometrically thin standard disk
around a black hole. This corona is fed by matter from the thin disk
which evaporates from the cool layers underneath.
An equilibrium establishes between the cool accretion stream and the
hot flow. In the
inner region evaporation becomes so efficient that
at low accretion rates all matter flows 
via the corona and proceeds towards the black hole as an
advection-dominated accretion flow (ADAF). We investigate this
transition for accretion disks around stellar black holes.
We derive a set of four ordinary differential equations which describe
the relevant processes and solve them numerically. The
evaporation efficiency has a maximum value where the
advection-dominated structure of the corona (at large distances from
the black hole) changes to a radiation-dominated structure (at small
distances). This maximum can be related to the observed spectral
transitions in black hole X-ray binaries. Our solutions of coronal
structure are valid for both, accretion in galactic X-ray transients
and in AGN.

\keywords{accretion disks -- black hole physics --  X-rays: stars
  -- stars: individual: Cygnus X-1, Nova Muscae 1991 -- galaxies: nuclei}
\end{abstract}

% 6 figures meyer.F1,....meyer.F6

\section{Introduction}
Recently a wealth of X-ray observations has provided more detailed
information about spectra of X-ray sources and short-time variability.
This concerns soft X-ray transients as good examples of stellar accreting
black holes, high-mass X-ray binaries with a black hole primary star
and also Active Galactic Nuclei. The new observations put more
stringent constraints on the modelling of accretion processes.
In this connection the change from accretion in a
standard cool disk to a hot coronal flow is an important feature.

Different solutions of accretion on to compact objects had been
investigated in the past. An account can be found in the review
by Narayan et al. (1998). The description of an optically thick,
geometrically thin disk, now often called ``standard thin disk'',
goes back to work of Shakura \& Sunyaev (1973), Novikov \& Thorne
(1973), Lynden-Bell \& Pringle (1974) and others. A much hotter
self-consistent solution of optically thin gas was discovered by
Shapiro, Lightman \& Eardley (1976), but found to be thermally
unstable by Piran (1978). A solution for even super-Eddington
accretion rates in which the large optical depth of the inflowing gas
prevents radiation to escape was suggested by Katz (1977), Begelman (1978),
Abramowicz et al. (1988) and others. This is an optically thick
advection-dominated accretion flow. The fourth solution concerns
accretion at low sub-Eddington rate (proposed by Shapiro et al. 1976,
Ichimaru 1977 and Rees et al. 1982). Detailed theoretical work on this
solution by several authors has proven that this optically thin regime
of solutions for low accretion rates is relevant for accretion in many
cases (Narayan \& Yi 1994, 1995a,b, Abramowicz, Chen \& Kato et
al. 1995, Chen 1995; Chen et al. 1995, 1997; Honma 1996; Narayan, Kato
\& Homna 1997a; Nakamura et al. 1996, 1997).

The model of an optically thin advection-dominated flow (ADAF) allows
to explain the accretion in soft X-ray transients during quiescence.
Assuming the old picture of a standard thin accretion disk reaching inward
towards the black hole the amount of X-ray flux is clearly inconsistent with
the accretion rate estimated from the outer optically thick disk
(McClintock et al. 1995). An interior ADAF allows to understand
the observations but requires the change from a standard thin disk to
the hot flow at
a certain transition radius. The black hole X-ray transients are a
laboratory for accretion on to black holes. The physics of the accretion
processes around supermassive black holes in AGN is similar, which
makes the investigation of ADAF even more important and fruitful.

The change from a geometrically thin disk to the hot flow is
an essential feature in this overall picture of accretion. The
``strong ADAF principle'', formulated by Narayan \& Yi (1995b),
suggests that, whenever the accreting gas has a choice between a
standard thin disk and an ADAF, the ADAF configuration is chosen.
This rule provides a transition radius $r_rm{\rm{tr}}$  for a
given accretion rate (see Fig. 6b in Esin et al. 1997). In a different
approach Honma (1996) considered the interface of a hot optically
thin disk with a  cool standard thin disk applying radial diffusive
transport of heat.

Here we present a model of a corona above a thin disk in order to
determine the inner edge of the thin disk, inside of which only a hot
optically thin flow exists. The
co-existence of a cool disk and a corona had been investigated for 
disks in dwarf nova systems, where features otherwise not understood
as for example the X-rays observed in quiescence can then be
explained (Meyer \& Meyer-Hofmeister 1994, Liu et al. 1995). The
situation is the same in disks around black holes (Meyer 1999), both
stellar and supermassive.

The physical two-dimensional situation would demand the
solution of a set of partial differential equations in radial distance
$r$ and vertical height $z$. In particular a sonic transition would require
treatment of a free boundary condition on an extended surface.
In Sect. 2 we show how we derive a simplified set of four ordinary
differential equations which describe the physics of the corona above the thin
disk. The interaction of corona and thin disk results in evaporation
of matter from the thin disk . In Sect. 3 we discuss the general
features of this evaporation process.  
The results of our computations are  presented in Sect. 4. 
The main result of this model is the efficiency of evaporation and with
this the location where the thin disk has completely changed to a hot
coronal flow (Sect. 5). In Sect. 6 we point out the invariance of our results
for different black hole masses. Sects. 7 and 8 concern
the validity of the assumptions made in our computations and the importance
of synchrotron and Compton processes. In Sect. 9 the comparison with
observations follows, in Sect. 10 a discussion of further effects.
In Sect. 11 we compare with the model of Honma (1996). Sect. 12
contains our conclusions.     

\section{The equations}
In this section we derive the equations that describe the corona above
the inner rim of the accretion disk. For this we will average over
an inner ring area and thereby reduce the set of partial differential
equations to a system of ordinary differential equations with respect
to height. The solutions then become a function of the radial coordinate of
the inner rim.

We start with the five equations of viscous hydrodynamics,
conservation of mass
\begin{equation}
\frac{\partial \rho}{\partial t} + \nabla \cdot \left(\rho
{\bf{v}}\right) = 0,
\end{equation} 
the equations of motion
\begin{equation}
\rho \left[\frac{\partial{\bf v}}{\partial t} + \left({\bf v}
\cdot {\bf \nabla} \right)\bf{v}\right] = - \nabla P + \nabla
\cdot  {\bf S}-
\rho {\bf \nabla} \mit\Phi,
\end{equation}
and the first law of thermodynamics
\begin{equation}
\rho \left[\frac{\partial U}{\partial t} + \left({\bf v}
\cdot {\bf \nabla} U \right)\right] = - P({\bf \nabla} \cdot{\bf
v}) + ({\bf S} \cdot \nabla) \cdot{\bf v} + \dot{q}, 
\end{equation}
supplemented by the equation of state for an ideal gas

\begin{equation}
P= \frac{\Re}{\mu} T \rho, 
\end{equation}
\begin{equation}
U = \frac{1}{\gamma-1} \frac{P}{\rho},
\end{equation}  
\begin{equation}
\frac{\Re}{\mu}T = V^2_s.
\end{equation}

Here $\rho, P$ and $T$ are density, pressure and temperature, ${\bf v}$ is
the velocity vector, ${\bf S}$ the viscous stress tensor, $\mit\Phi$ the
gravitational potential, $U$ the internal energy, $\Re$ the gas
constant, $\mu$ the molecular weight. We take $\mu= 0.62$ for a
fully ionized gas of cosmic abundances. $V_{\rm s}$ is the isothermal speed of
sound, $\gamma$ the ratio of the specific heats, = $\frac{5}{3}$,
for a mono-atomic gas. For ${\bf S}$ we take the form appropriate for a
mono-atomic gas with Cartesian components
\begin{equation}
\sigma_{ik} =
 \mu_{\rm v}\left[\frac{\partial v_i}{\partial x_k} + \frac
{\partial v_k}{\partial x_i}-\frac{2}{3} \delta_{ik} \frac{\partial
v_l}{\partial x_l}\right] 
\end{equation}
(Sommerfeld 1949). For the (non-molecular) viscosity we
use an $\alpha$-parameterization (Shakura \& Sunyaev 1973, Novikov
\& Thorne 1973)
\begin{equation}
\mu_{\rm v}=\frac{2}{3} \alpha P/\mit\Omega,
\end{equation}
$\mit\Omega$ rotational frequency. $\dot q$ is the rate of heating per
unit volume. We write it in two parts, 

\[\dot{q} =\dot{q}_{\rm{rad}} + \dot{q}_{\rm{cond}}\] 
where $\dot q_{\rm{rad}}$ is the gain or loss of heat by radiation and
$\dot{q}_{\rm{cond}}$ that by heat conduction

\begin{equation}
\dot{q}_{\rm{cond}}= - \bf{\nabla} \cdot \bf{F_{\rm {c}}}
\end{equation}
For $\dot q_{\rm{rad}}$ we take here optically
thin thermal radiation loss according to Fig. 1 of Raymond et al. (1976).
\begin{equation}
\dot{q}_{\rm{rad}}= - n_{\rm e}\,n\,\Lambda(T)
\end{equation}
where $n_{\rm e}$ and $n$ are electron and ion particle densities.
For the heat flux $\bf{F_{\rm{c}}}$ we take 
\begin{equation}
{\bf F}_{\rm{c}} = -\kappa_0 T^{5/2} {\bf\nabla}T ;
\end{equation}
(Spitzer 1962) for a fully ionized plasma and use the value 
\[
\kappa_0 = 10^{-6}
\frac{\rm{g\,cm}}{\rm{s^3K^{7/2}}} 
\]\
neglecting a weak logarithmic dependence on electron density and
temperature.

Taking the scalar product of Eq.(2) with $\bf{v}$, adding to Eq.(3),
and using
Eq.(1) we obtain the equation of conservation of energy
\begin{eqnarray}
\lefteqn{
\frac{\partial }{\partial t}\left[\rho (U +  \frac{v^2}{2} + 
\mit\Phi)\right]+
{\bf\nabla} \cdot \bigg[{\bf {v}}\bigg(\rho U +
\rho \frac{v^2}{2} + \rho \mit\Phi+P\bigg)
}\nonumber \\
& &
-({\bf
S\cdot\bf{v}})+{\bf {F_{\rm c}}} \bigg]= \dot{q}_{rad}
\end {eqnarray}

We introduce cylindrical coordinates, $r,\varphi,z$ with the $z$-axis
perpendicular to the midplane of the disk and through the central
accretor, and use the vector correspondence to the Cartesian advection
term

\begin{equation}
({\bf v} \cdot {\bf \nabla}){\bf v} =
{\bf \nabla} \frac{v^2}{2} - {\bf v} \times
({\bf \nabla} \times {\bf v}),  
\end{equation}
where $v^2\equiv ({\bf v} \cdot {\bf v})$. The (Newtonian) potential is then 
\begin{equation}
\Phi = - \frac{GM}{(r^2+z^2)^{1/2}} 
\end{equation}
with $G$ the gravitational constant and $M$ the central mass.

We consider from now on stationary, azimuthally symmetric flows.

In the following subsections we derive values for the rotational
velocity $v_\varphi$ and the radial diffusive velocity $v_r$, and the three
remaining ordinary differential equations for the conservation of
mass, the vertical dynamic equilibrium, and the energy flow in the
corona above the inner zone of the accretion disk.

For this we choose a region between inner disk radius $r_1$ and
radius $r_2$ such that 
\begin{equation}
\pi(r_2^2-r_1^2) =\pi r^2, 
\end{equation}
where $r$ is an area weighted mean radius between $r_1$ and $r_2$.
We apply therefore, at each radius r, the integration
bounds $r_1$=0.72\,$r$ and  $r_2$=1.23\,$r$. The coronal solution
depends on the value of $r$ as a measure of the distance from the accretor.

We estimate the relative order of magnitude of the various terms in
our equations and drop non-dominant terms. For these estimates we use
relations that result from our solutions which are therefore
self-consistent.

All our solutions have 

\begin{equation}
\frac{{\Re/\mu} T r }{GM}=\frac{V_{\rm {s}}^2r}{GM}\leq 1/8,
\end{equation}
see Fig. 3. 
They are subvirial and therefore like ``slim'' accretion disks
(Abramowicz et al. 1988) with cylindrical  symmetry in
their (main) part in contrast to the more spherical symmetry of ADAF
solutions. As a result

\begin{equation}
\left|\frac{v_{r}}{v_{\varphi}}\right| \: \approx \alpha
\frac{V_{\rm{s}}^2}{v_{\varphi}^2}\;\le
\frac{\alpha}{8},
\end{equation}
where we have replaced the azimuthal velocity by its Kepler value
which is a good approximation in the main body of our solution, see
the remark follwing Eq. 25 below. 
A wind from such a corona reaches escape speed at
large height leading to
\begin{equation} 
v_{z}\leq V_{\rm s}\frac {z}{2r}, 
\end{equation}
see Fig. 2. Where in the following we neglect radial derivatives of
velocity components we use the order of magnitude relation
\begin{equation}
\frac{\partial v}{\partial r} =O(\frac{v}{r}).
\end{equation}

\subsection{The $r$-component of the equation of motion}

This equation yields the rotational velocity. 

The radial component of the equation of motion becomes 

\begin{eqnarray}
\rho\left(v_r\frac{\partial v_r}{\partial r}+v_z \frac{\partial
v_r}{\partial z}-\frac{v_\varphi^2}{r}\right)
& =  & -\frac{\partial
P}{\partial r}-\rho \frac{GMr}{(r^2+z^2)^{3/2}} \nonumber\\  & & +
\frac{1}{r}\frac{\partial}{\partial r}(r\sigma_{rr})+\frac{\partial}
{\partial z} \sigma_{rz} 
\end{eqnarray}

In this equation the pressure term $\partial P/\partial r$ is small
compared to the gravitational term in regions $z \leq r/2$
dominant with respect to dynamics and energetics of the solutions (see
Fig. 2). Taking $\partial P/ \partial r$ =$O(P/r)$ we
estimate

\begin{eqnarray}
\bigg|\frac{\partial P}{\partial r}\bigg| \bigg/
\frac{\rho GMr}{(r^2+z^2)^{3/2}}
& &\approx
 \frac{{V_{\rm{s}}^2}r}{GM} (1+\frac{z^2}{r^2})^{3/2} \nonumber \\
& & 
\leq \frac{1}{8}(1+\frac{z^2}{r^2})^{3/2} \leq 0.18
\end{eqnarray}

This shows a relative unimportance of radial pressure support in these
 coronae (in contrast to virialized ADAF solutions, Narayan \& Yi
(1995b)).
At a large height $z=r$ this ratio has increased to 0.35. But we note
that with the topology of Fig. 1 that at that height the radial
pressure gradient must have significantly decreased because of the
axial geometry. 
Therefore to first approximation we can neglect the pressure term.

The viscous terms
\begin{equation}
 \sigma_{rr} =\mu_{\rm v}\big(2\frac{\partial v_{r}}{\partial r}
-\frac{2}{3}\nabla\cdot{\bf{v}}\big),\;\;
 \sigma_{rz}=\mu_{\rm v}
\big(\frac{\partial v_{r}}{\partial z} +\frac{\partial v_{z}}{\partial r}
\big)
\end{equation}
contribute derivatives of $r$- and $z$-components of the velocity.
Their contributions, evaluated with Eqs. (16) to (19), are small compared to
\begin{equation}
\frac{\partial P}{\partial r}
\approx O(P/r)
\end{equation}
by at least a factor $V_{s}/v_{\varphi}$.
They as well as the first two
terms on the left hand side of Eq. (20) are negligible. 

This leaves the balance between centrifugal and radial gravitational
force

\begin{equation}
v_{\varphi}^2=\frac{GM}{r}(1+\frac{z^2}{r^2})^{-3/2}
\end{equation}
\begin{equation}
\mit{\Omega} =
{\sqrt{\frac{GM}{r^3}}}(1+\frac{z^2}{r^2})^{-3/4} 
\end{equation}

In the main body of the coronal solution the Kepler values
(i.e. at $z$=0) are a good approximation to $v_\varphi$ and $\mit\Omega$.
In our calculation we take the Kepler values throughout as a
computational convenience. It overestimates the frictional work and
underestimates the release of rotational energy at large height. These
two effects partially compensate each other. At $z\leq r/2$ the sum of
these effects deviates from the
accurate value by less than $7\%$. At $z=r$ it is 20$\%$. But at this height
the total further contribution of both energy terms already is
negligibly small (due to the exponential decrease of pressure and
density with height, see Fig. 2) and isothermality of the solution
completely dominates. This simplification has thus only a small effect
on the accuracy of the solution.

\subsection{The $\varphi$-component of the equation of motion} 

The conservation of the $z$-component of angular momentum
yields the $r$-component of the velocity.
 
Multiplication of the equation of motion by $r\,{\bf u}_{\varphi}
({\bf u}_r,{\bf u}_{\varphi},{\bf u}_z$ unit vectors in  $r$,
$\varphi$ and $z$ direction) yields 

\begin{eqnarray}
\rho {\bf v} \cdot {\bf \nabla}(r v_{\varphi}) & 
= & r{\bf u}_{\varphi}\cdot({\bf\nabla} \cdot S) \nonumber \\&
= & \nabla \cdot({\bf S} \cdot r {\bf u}_{\varphi})-({\bf
S}\cdot\nabla)r{\bf u}_\varphi 
\end{eqnarray}
Using (1) (for $\frac{\partial}{\partial t}$ =0)
\[ {\bf\nabla} \cdot \rho {\bf v}=0,  \]
and
\[ {\bf\nabla}r{\bf u}_{\varphi} =
{\bf u}_r {\bf u}_{\varphi}-{\bf u}_{\varphi}{\bf u}_r, \]
(the right side understood as the dyadic product)
and
\[
\sigma_{r \varphi}=\sigma_{\varphi r}, \]
this results in the equation of conservation of $z$-angular momentum
\begin{equation}
{\bf\nabla} \cdot (\rho {\bf v} r {v}_\varphi - r {\bf
u}_\varphi \cdot {\bf S}) = 0.
\end{equation}

We integrate over a cylinder with radius $r_1$ and obtain

\begin{equation}
2\pi {r_1}^2 \int^ \infty_0 \rho {v}_r {v}_\varphi dz =
2\pi {r_1}^2 \int^ \infty_0 \sigma_{r \varphi} dz.
\end{equation}

Here it is assumed that no significant angular momentum is accreted at
the center of the compact object or lost through the boundary
$z=\infty$.
For accretion and jet angular momentum loss from the most interior
parts of the accretion region, such losses would be small by order of 
$\sqrt{R/r_1}$ where $R$ is the radius of the inner accretion region,
e.g. $R\approx 3r_{\rm S}$ for the Schwarzschild radius $r_{\rm S}$ of
a central black hole ($r_{\rm S}=2GM/c^2$).

We replace the global balance of angular momentum flow of Eq. (28) by
its local equivalent and obtain the radial diffusive velocity

\begin{equation}
 {v}_r = \frac{\sigma_{r \varphi}}{\rho{v}_\varphi}
= -\alpha\frac{V_s^2}{{v}_\varphi}
\end{equation}
with $\sigma_{r \varphi}= \mu_{\rm v}(\partial v_{\varphi}/\partial r
-v_{\varphi}/r) =-\alpha P$. This is the standard bulk radial velocity
formula for accretion disks. It neglects the modification by
$z$-dependent secondary flows superimposed on the main drift of
material (see Kley \& Lin 1992).
Such secondary flows may slightly modify the local vertical structure
but will hardly affect the general structure of the solutions.

\subsection{The $z$-component of the equation of motion}

The vertical component of the equation of motion Eq. (2) yields the
vertical dynamic equilibrium.

The vertical component becomes 

\begin{eqnarray}
\lefteqn{\rho ({v}_r\frac{\partial{v}_z}{\partial r} +
{v}_z \frac{\partial{v}_z}{\partial z}) = -\frac{\partial P}
{\partial z}-  \rho \frac{GMz}{(r^2+z^2)^{3/2}}} \nonumber \\
& & +\frac{1}{r} \frac{\partial}{\partial r}(r \sigma_{r z})+
\frac{\partial}{\partial z}\sigma_{zz}. 
\end{eqnarray}

Using approximate isothermal pressure layering (see Fig. 2)
\begin{equation}
P=P_0\,e^{-(z/h)^2}
\end{equation}
with $h\approx\frac{1}{2}r$ (see Fig. 3) and the estimates of Eqs. (16)
to (19) we estimate

\begin{equation}
\bigg| \rho v_r \frac{\partial v_z}{\partial r}\bigg|  \bigg/
\bigg| \frac{\partial P}{\partial z}\bigg|  \leq 0.02\alpha
\end{equation}
to be negligible. Further, with 
\begin{equation}
\sigma_{rz}=\mu_{\rm v}\left( \frac{\partial v_r}{\partial z}
+\frac{\partial v_z}{\partial r}\right),
\sigma_{zz}=\mu_{\rm v}\left(2 \frac{\partial v_z}{\partial z}
-\frac{2}{3}\left(\nabla\cdot {\bf v}\right)\right)
\end{equation}
the largest contribution of all viscous terms in Eq. (30) appears in 
$\partial\sigma_{zz}/\partial z)$ from the $z$-derivative of $\mu_{\rm v}$
(that is of $P$, see Eq. (8) times the derivative $\partial
v_z/\partial z$. This term is of order $\leq$0.16$\alpha\approx$0.05
and negligible. In the remaining dominant terms only $z$-derivatives
appear. Averaging over the inner zone we obtain

\begin{equation}
\rho v_z\frac{dv_z}{dz}=-\frac{dP}{dz} -\rho
\frac{GMz}{(r^2+z^2)^{3/2}}.
\end{equation}

\subsection{Mass conservation}

The equation of mass conservation we integrate over the radial zone
between $r_1$ and  $r_2$, of area $\pi r^2$  and obtain

\begin{equation}
\int^{r_2}_{r_1} 2 \pi r \frac{\partial}{\partial
z}(\rho{v}_z)dr = (2\pi r\rho{v_r})_1
-(2\pi r\rho{v_r})_2.
\end{equation}

When, as we assume, $r\,\rho\,v_r$ drops off with radius strongly
(e.g. $\rho \sim r^{-3}$ in the scaled solutions, Liu et al. 1995) the
difference on the right hand side is dominated by the first term, and
its value will be slightly smaller than this term. We approximately
take this into account by replacing the term at $r_1$ by its value
at the somewhat larger mean radius $r$.
We replace the left hand side by the area times a
mean value. This leads to the approximation
\begin{equation}
\frac{d}{dz} (\rho{v}_z) =
\frac{2}{r} \rho{v}_r 
\end{equation}
where $v_r$ is the radial flow velocity (Eq. 29). At large height,
however, we have to take into
account the flaring of the channel as the ascending gas changes from
a cylindrical rise to spherical expansion (see Fig. 1). How can we
account for the additional attenuation of the vertical flux density by
this change of geometry?

We proceed in the following way. Approximately the cross section of
the vertical flow channel can be taken as proportional to
$ 1+z^2/r^2$. Without sidewise diffusive loss, mass
conservation would then yield 

\begin{equation}
\frac{d}{dz}\left[\left(1 + \frac{z^2}{r^2}\right)\rho{v}_
z\right] = 0
\end{equation}
if we take the center of the expanding flux tube in the vertical
direction. Per unit cross section the resulting conservation equation
is then

\begin{eqnarray}
%\lefteqn{\frac{1}{1+\frac{z^2}{r^2}} \frac{d}{dz}\left[\left(1 +
%\frac{z^2}{r^2}\right) \rho{v}_z\right] =
%\frac{d}{dz}\left(\rho{v}_z\right) + \frac{2z}{r^2+z^2} \rho{v}_z
% \nonumber} \\
%& &= 0
\frac{1}{1+\frac{z^2}{r^2}} \frac{d}{dz}\left[\left(1 +
\frac{z^2}{r^2}\right) \rho{v}_z \right] & = &
\frac{d}{dz}\bigg(\rho{v}_z\bigg) + \frac{2z}{r^2+z^2} \rho{v}_z
\nonumber \\
& = & 0
\end{eqnarray}

The gas in a cylindrical column thus experiences an additional
geometry-related radial divergence
\[ -\frac{2z}{(r^2+z^2)}\;\rho{v}_z \]
Adding the two radial flow components we obtain the final form
\begin{equation}
\frac{d}{dz}(\rho{v}_z)= \frac{2}{r}\rho{v}_r -
\frac{2z}{r^2+z^2} \rho{v}_z   
\end{equation}

We note that the diffusive part in this equation becomes unimportant
at large height where the diffusive description for the flaring flux
tube
becomes problematic.

\subsection{Energy conservation}
We first derive the frictional flux terms. With Eqs. (22) and (13)
we obtain

\begin{equation}
({\bf v} \cdot {\bf S}) =\mu_{\rm v} \bigg[ 
{\bf \nabla} v^2 - {\bf v} \times
({\bf \nabla} \times {\bf v}) -\frac{2}{3} {\bf v}( \nabla \cdot
{\bf v})\bigg]
\end{equation}
and its $r$-component 

\begin{eqnarray}
{\bf u}_r \cdot({\bf v}\cdot {\bf S})&=&
\mu_{\rm v} \bigg[ \frac{\partial v^2}{\partial r} -v_\varphi
\frac{\partial v_\varphi}{\partial r} -\frac{{v_\varphi}^2}{r}
 \nonumber\\
 & & +v_z\left(\frac{\partial v_r}{\partial z}-\frac{\partial v_z}
{\partial r}\right)
 -\frac{2}{3} v_r \nabla \cdot {\bf v} \bigg]
\end{eqnarray}

Here all terms are negligible except those containing $v_\varphi$, by
use of Eqs. (16) to (19). Making use of Eqs. (8) and (19) we obtain 
\begin{equation}
{\bf u}_r \cdot({\bf v} \cdot {\bf S}) = -\alpha P v_\varphi.
\end{equation}
Likewise, only keeping non-negligible $v_\varphi-$ terms, the $z$-component 
is

\begin{eqnarray}
{\bf u}_z \cdot({\bf v}\cdot {\bf S})=
\mu_{\rm v} \bigg[ \frac{\partial v^2}{\partial z}
- v_r\left(\frac{\partial v_r}{\partial z}-\frac{\partial
v_z}{\partial r}\right) \nonumber\\
 -v_\varphi \frac{\partial v_\varphi}{\partial z}
 -\frac{2}{3} v_z \left(\nabla \cdot {\bf v}\right) \bigg]
=\frac{2}{3} \alpha P r \frac{\partial v_\varphi}{\partial z}.
\end{eqnarray}

The radial component of the divergence can be written, using (19) as

\begin{eqnarray}
\frac{1}{r}\frac{\partial}{\partial r}\big [r{\bf u}_r \cdot({\bf
v}\cdot {\bf S})\big]=
-\alpha P \mit\Omega \bigg(\frac{1}{2}+\frac{\partial \rm{ln}P}{\partial
\rm{ln}r}\bigg) = \frac{3}{2}\alpha P \mit\Omega.  
\end{eqnarray}

For $\partial \rm{ln} P /\partial \rm{ln} r$ we have here taken its estimated
value -2 at the characteristic height $z=h=r/2\;\;$ (see Fig. 2), which we
obtain from Eq. (31) with the scaling $P_0\sim r^{-4}$ that holds when we
compare solutions with different distances $r$ (see Liu et al. 1995).
The resulting term represents the dominant heating mechanism of the
corona by release of gravitational energy.

We now integrate over the area between $r_1$ and $r_2$ as in the
derivation of conservation of mass 

\begin{eqnarray}
\lefteqn{
\int^{r_2}_{r_1} 2 \pi r dr \bigg\{ \frac{\partial}{\partial z}
\bigg[{v}_z\left(\rho \frac{{v}^2}{2} +
\frac{\gamma}{\gamma-1}P + \rho{\mit\Phi} \right)
   } \nonumber
\\ & &
+(F_c)_z+\frac{2}{3}
\alpha P r \frac{\partial v_\varphi}{\partial z}\bigg] -\frac{3}{2}
\alpha P {\mit\Omega}-\dot{q}_{rad}\bigg\}
\\ & = & \bigg\{2\pi r \bigg[{v}_r\bigg(\rho\frac{v^2}{2}
+\frac{\gamma}{\gamma-1} P+\rho{\mit\Phi} \bigg) +(F_{\rm c})_r \bigg]
 \bigg\}_{r_1} 
\nonumber
\\ & &
-\bigg\{ \ldots \bigg\}_{r_2}\nonumber
\end{eqnarray}

On the right hand side of this equation we again replace the difference
of the terms taken at $r_1$ and $r_2$ by the value taken at the
intermediate value $r$, because the value at $r_2$ will only be a
fraction  of the value at $r_1$ due to the radial drop off of all the
quantities. If we take $v_r$ as the diffusive velocity (Eq. (29) we
must still include the divergence of the vertical flow due to the
flaring geometry. We use the same procedure as in the preceding
Sect. 2.4 to derive the corresponding term and add it on the right
hand side. Again, on the left hand side of the equation we write the
integral as area times a mean value and obtain the energy equation in
its final form
 
\begin{eqnarray}
\lefteqn{\frac{d}{dz}\bigg[{v}_z\left(\rho\frac{v^2}{2}+\frac{\gamma}
{\gamma-1}P+\rho\mit\Phi\right)
 + \frac{2}{3}
\alpha P r \frac{dv_{\varphi}}{dz}  +(F_{\rm c})_z \bigg]=}
 \nonumber \\
& &
\frac{3}{2}\alpha P{\mit\Omega} +\dot{q}_{rad}
 \nonumber \\
& & +\frac{2}{r} \bigg[{v}_r
\bigg(\rho\frac{{v}^2}{2}
+\frac{\gamma}{\gamma-1} P+\rho\mit\Phi\bigg) 
+(F_{\rm c})_r\bigg]
 \nonumber \\
& & - \frac{2z}{r^2+z^2}\bigg[{v}_z\bigg(\rho\frac{{v}^2}{2}
+\frac{\gamma}{\gamma-1} P +\rho\mit\Phi\bigg)
\nonumber \\
& &
+ \frac{2}{3}\alpha Pr\frac{dv_\varphi}{dz}+ (F_{\rm c})_z \bigg].
\end{eqnarray}

Had we chosen to integrate the radial divergence
$\frac{1}{r}\frac{\partial}{\partial r} \big[r{\bf u}_r \cdot ({\bf v} \cdot
{\bf S})\big]$  underneath the integral of Eq. (45) instead
of evaluating this term as in Eq. (44) the last equation would have the
factor 2 instead of $\frac{3}{2}$. This difference arises from the
different assumptions about radial variation of the
quantities, implicit in the two approaches. It shows a
natural limit of numerical accuracy of any such approach.

The lateral inflow of heat by $(F_{\rm c})_r$ adds to the frictional
release of energy in the coronal
column. We show that the relative size of the contribution is small
and we will take $(F_c{\rm})_r$=0 in our computations. We assume that
the temperature at low $z/r$ varies like the virial temperature,
$T\sim 1/r$, and that at $z\geq r$ the radial temperature gradient becomes
small as the topology changes and the corona fills the full
cylindrical region, with $\frac{\partial T}{\partial r}=0$ on the axis
of symmetry. With these assumptions we take
\begin{equation}
\frac{\partial T}{\partial r}=
\frac{T}{r}\left(1-\frac{z^2}{r^2}\right).
\end{equation}
The ratio of total radial conductive heat input $(\dot{Q}_{\rm c}
)_r$ 
to the total frictional energy input $\dot{Q}_{\rm f}$ is then
evaluated as
\begin{eqnarray}
\lefteqn{\frac{(\dot{Q}_{\rm c}
)_r}{\dot{Q}_{\rm f}}
  } \nonumber\\ & &
=\int^{z_0}_{0}\frac{1}{r}\frac{\partial}{\partial
r} \left[r(F_c)_r\right] dz \bigg/ \int^{z_0}_{0}\frac{3}{2} \alpha P
{\mit\Omega}dz=0.22, 
\end{eqnarray}

It is independent of the value of $r$ as long as we are in the
``advection-dominated'' range $r \geq r(\dot{M}_{\rm {crit}})$ (see
Fig.3). This follows when one uses the scaling of $P$, $T$ and
$\mit\Omega$ derived by Liu et al. (1995) (see also values $P_0$ in Table 1).
In the ``conduction-dominated'' corona $r \leq r(\dot{M}_{\rm
{crit}})$ the temperature becomes
constant and the ratio becomes zero. The contribution of 
$(F_{\rm c})_r$ can therefore be neglected. The remaining
$z-$component of Eq. (11) we write as

\begin{equation}
F_{\rm c}\equiv (F_{\rm c})_z = -\kappa_0 T^{5/2}\frac{dT}{dz} 
\end{equation}

\subsection {The system of ordinary differential equations with
boundary conditions}

From the five partial differential equations (1) to (3) we have
now derived three ordinary differential equations (34), (39) and
(46) (where we take $(F_{\rm c})_r=0$) which describe vertical dynamical
equilibrium and conservation of mass and energy, plus the two
expressions
for $v_{\varphi}$ and $v_r$ (Eqs. (24) and (29)). Our fourth ordinary
differential equation is the dependence of the heat flux on the
temperature gradient, Eq. (49).
As the 4 dependent variables we have temperature $T$, pressure $P$, mass flow
$\dot m=\rho\,v_z$ and heat flux $F_{\rm c}$. The 4 boundary
conditions are: 

(1) No pressure at infinity (i.e. no artificial confinement). This
requires sound transition at some height $z=z_1$ (free boundary
problem)
\begin{equation}
v_z=V_{\rm s} \quad\mbox{at}\quad z=z_1.
\end{equation}

(2) No influx of heat from infinity. We neglect a small outward
conductive heat flux induced by the temperature gradient in the
expanding wind and require
\begin{equation}
F_{\rm c}=0  \quad\mbox{at}\quad z=z_1.
\end{equation}

(3) Chromosphere temperature at the bottom $z=z_0$
\begin{equation}
T=T_{\rm{eff}} \quad\mbox{at}\quad z=z_0.
\end{equation}

(4) No heat inflow at the bottom  
\begin{equation}
F_{\rm c}=0  \quad\mbox{at}\quad z=z_0.
\end{equation}
This approximates the condition
that the chromospheric temperature is kept up by other processes and
cannot conduct any sizeable thermal heat flow. 

The consequence of conditions (3) and (4) is that the conductive flux
that comes down at a certain level above the boundary must be radiated
away. Since the advective cooling by the rising mass flow is small (in
our solutions at $T^{6.5}K$ only $\frac{1}{10}$) this reduces to the
balance between thermal conductive flux and radiation and establishes
the temperature profile as the one studied by Shmeleva \&
Syrovatskii (1973) for the solar corona. In these lowest layers the
pressure is nearly constant and the temperature rises steeply. Liu et
al. (1995) recalculated the temperature profile using
the more recent cooling function of Raymond et al. (1976) and we used the
improved version. This allows to replace the conditions (3) and (4) by
\begin{equation}
T=10^{6.5}\rm K \quad\mbox{and}\quad
  F_{\rm c}=-2.73\,10^6\,P \quad\mbox{at}\quad z=z_0 
\end{equation}
The relation between $P$ and $ F_{\rm c}$ is practically independent of the
value of the chromospheric temperature as long as it is small compared
to $T_0=10^{6.5}$K. 

These equations and boundary conditions describe a model of the
corona-disk interaction for the innermost region of the thin accretion
disk where evaporation is most efficient.

\section{General features of evaporation}

The equations derived in the last section describe the physics of 
a hot corona in equilibrium with a thin cool disk underneath. 
Aspects of this model had already been investigated for accretion
on to a white dwarf (Meyer \& Meyer-Hofmeister 1994, Liu et al. 1995,
for application to WZ Sge stars and X-ray transients see also
Mineshige et al. 1998).
The picture that arises from the processes involved is the following.
Heat released by friction in the corona flows down into cooler and
denser transition layers. There it
is radiated away if the density is sufficiently high. If the density is
too low, cool matter is heated up and evaporated into the corona until
an equilibrium density is established. Mass drained from the corona by
an inward drift is steadily replaced by mass evaporating from the thin disk as
the system reaches a stationary state. This evaporation rate increases
steeply with decreasing distance from the central compact star.
When the evaporation rate exceeds the mass flow rate in the cool disk, a hole 
forms up to the distance where both rates are equal. Inside this 
transition radius $r_{\rm{tr}}$ only a hot coronal flow exists.

The mass flow in the corona is similar to that in the thin disk. The
matter has still its angular momentum, and differential (Kepler-like)
rotation causes friction. The main part of the coronal gas flows towards the
center directly. For conservation of angular momentum a small part
flows outward in the coronal layer and condensates in the outer cool
disk. Since the temperature is very high the corona is geometrically
much thicker than the disk underneath.
Therefore processes as sidewise energy transport and downward
heat conduction are important for the equilibrium between
corona and thin disk, together with radiation and wind loss.

We model the equilibrium between corona and thin disk in a simplified
way using a one-zone model as described in Sect. 2. This is possible
since the evaporation process is concentrated
near the inner edge of the thin disk and we can choose a
representative radial region from $r_1$ to $r_2$ such that the main
interaction between cool disk and corona occurs there and evaporation
further outward is not important. In Fig. 1 we show the flow pattern
as derived from analysis and simplified modelling for the case that a
hole in the thin disk exists. There are three regimes. 
(1) Near the inner edge of the thin disk the gas flows towards the
black hole. (2) At larger $r$ wind loss is important taking away about
20\% of the total matter inflow. (3) At even larger distances some matter
flows outward in the corona as a consequence of conservation of angular
momentum. (One might compare this with the flow in a ``free'' thin
disk without the tidal forces acting in a binary. In such a disk
matter flows inward in the inner region and outward in the outer
region, with conservation of the total mass and angular momentum
(Pringle 1981)). 

\begin{figure}[ht]
\includegraphics[width=8.8cm]{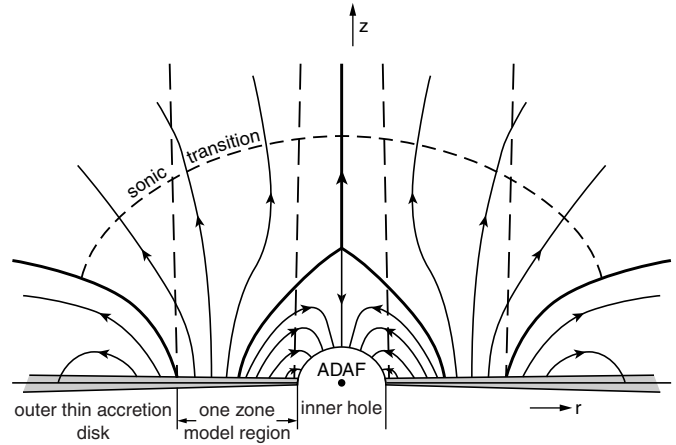}
\caption{Coronal mass flow pattern (schematic).
Convective circulations will be superimposed.}
\end{figure}

\section{Results from computations}

\subsection{Computer code technique}
We solve the differential equations (34), (39), (46), and (49) using
the Runge-Kutta method integrating from the lower boundary at an
assumed height $z_0\ll r$ above
the midplane to the upper boundary at $z_1$. The value $z_1$
is determined by our computations as the height where sound transition
occurs (free boundary problem). The boundary conditions are given in
Eqs. (50), (51) and (54). We require that the energy flux at the upper
boundary
is zero but are content in our computations if it is very small,
$|F_c| \leq 0.005 \cdot |(F_c)_{\rm{max}}|$.

We have to vary $P_0$ and ${\dot m_0}$,
trial values for $P$ and ${\dot m}$ ($\dot m=\rho v_z$) at $z=z_0$,
to find the coronal structure which fulfills the upper
boundary conditions. The conditions,
sound transition and zero heat flux, are met only for a unique pair of
values for pressure and mass flow at the bottom. For other values
sound transition is not reached and/or the heat flux is not
negligible (this is the same situation as in coronae of cataclysmic
variable disks,
compare Fig. 2 of Meyer \& Meyer-Hofmeister 1994). Numerically, the smaller the
distance $r$ from the compact object the more accurately the pressure
value has to be determined to obtain a consistent solution.

For our computations we put the lower boundary
at about $z_0=2\,10^6$cm above the midplane, but for $r \leq 3\,10^8$
cm at
correspondingly smaller values $z_0$. In most cases the true height
of the thin disk is less than the chosen value $z_0$. We carried out test
computations to check the influence of assuming different values $z_0$
on the coronal structure. As expected the changes were found to be very small.

$P_0$ and $(F_{\rm c})_0$ are nearly independent of the value of the
chromospheric temperature as long as it is small compared to
$T_0=10^{6.5}$K.

Since our results also concern
accretion in AGN we have confirmed the validity of our
assumption for the location of the lower boundary for this case also.

\subsection{Coronal structure}

We took 6$M_\odot$ as our standard parameter value to model the
evaporation in accretion disks around stellar black holes. We compute the
structure of the corona for distances of the outer cool disk from the black
hole ranging from large values to
small values of a few hundred Schwarzschild radii. For different
distances $r$ the structure in $z$ direction
is similar if considered as a function of $z/r$. As will be discussed
in Sect. 6 the coronal structure scales with Schwarzschild radius
and Eddington accretion rate.

\begin{figure}[h]
\includegraphics[width=7.5cm]{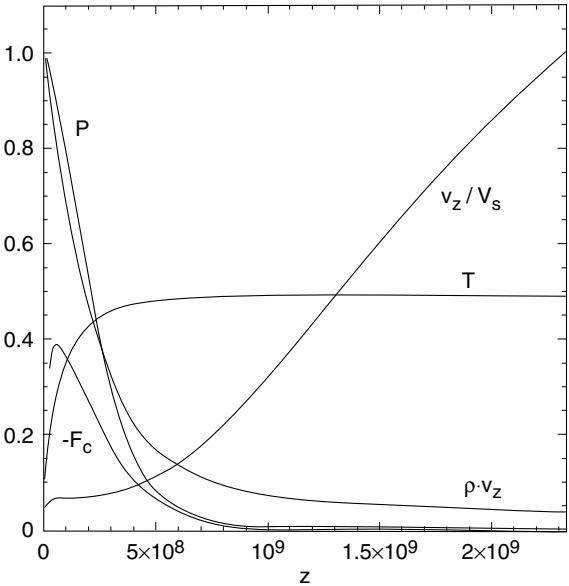}
\caption{Coronal structure at distance $r=10^{8.8}$cm from the 6$M_\odot$
black hole. Temperature $T$ is given in units of 0.2$ \cdot$virial
temperature, pressure P and vertical mass flow $\rho \cdot v_z$
($v_z$ vertical velocity) are scaled to their values at the lower
boundary. -$F_{\rm c}$ is the downward conductive heat flux, measured in
units of the frictional energy generation rate in a column of unit cross 
section. $V_{\rm s}$ is the sound velocity.}
\end{figure}

In Fig. 2 we show the structure at $r=10^{8.8}$cm close to where the
evaporation efficiency is highest for a 6$M_\odot$ black
hole. Pressure and density decrease significantly within a narrow
range above the cool disk. The downward conductive
heat flux -$F_{\rm c}$ therefore also has its maximum at a low height $z$, 
then rapidly decreases with height and is negliglibly small where sound
transition $ v_z/V_{\rm s}=1$ is reached, our boundary conditions.
This happens at 2.3\,$10^9$cm. The temperature is
practically constant in the higher layers of the corona, a consequence
of the high thermal conductivity.
The computed coronal structure at distance $r$ corresponds to the
situation that the cool disk underneath extends from this distance $r$
(inner edge) outwards. The derivation in Sect. 2 is based on the
dominant role of the innermost corona and our model
cannot be used to describe the coronal structure above the cool thin
disk farther out which will be affected by the dominant process
farther in.

These results can be compared to those of our earlier investigation of
evaporation in
cataclysmic variable disks. The coronal structure  was considered
there at distances
$10^{8.7}$ to $10^{9.7}$cm from the 1$M_\odot$ white dwarf  adequate
for a disk around a white dwarf. Due to the scaling
with the mass of the compact star the structure above the thin disk around
a 1$M_\odot$ white dwarf at distances log\,r 9.0 and 9.7 (shown in Liu et
al. 1995, Fig. 1a and 1b) corresponds to the structure at
distances $\rm {log}\, r/r_{\rm S}=3.53$ and 4.23 from a 6$M_\odot$ black hole.

\begin{table*}
\caption[]{\label{t:evapr-bh} Evaporation of the disk around 
 a black hole  of 6$M_\odot$}
\begin{center}
\begin{tabular}{cccccccc}
\hline\hline
\noalign{\smallskip}

$\log r$&$\log \dot m_0$& $\log P_0$&$ T$(K)&$\dot M=2\pi r^2\dot
m_0 (M_\odot /y)$&$\lambda$&$\log r/r_{\rm S}$&
$\dot M/{\dot M_{\rm Edd}}$\\
\noalign{\smallskip}\hline\noalign{\smallskip} 
10&-4.644&4.140&$8.00\times10^7$&$2.263\times10^{-10}$&0.211& 3.75 &
$1.712\times10^{-3}$\\
9.7&-3.646&5.302&$1.55\times10^8$&$5.657\times10^{-10}$&0.185& 3.45 &
$4.280\times10^{-3}$\\
9.3&-2.399&6.791&$3.70\times10^8$&$1.582\times10^{-9}$&0.146& 3.05 &
$1.197\times10^{-2}$\\
9.0&-1.586&7.825&$6.67\times10^8$&$2.583\times10^{-9}$&0.091& 2.75 &
$1.954\times10^{-2}$\\
8.9&-1.351&8.144&$7.95\times10^8$&$2.800\times10^{-9}$&0.066& 2.65 &
$2.118\times10^{-2}$\\
8.8&-1.143&8.445&$9.28\times10^8$&$2.852\times10^{-9}$&0.039& 2.55 &
$2.158\times10^{-2}$\\
8.7&-0.975&8.717&$1.04\times10^9$&$2.650\times10^{-9}$&0.016& 2.45 &
$2.005\times10^{-2}$\\
8.5&-0.824&9.112&$1.09\times10^9$&$1.493\times10^{-9}$&$\le 10^{-3}$& 2.25 &
$1.130\times10^{-2}$\\
8.3&-0.746&9.421&$1.05\times10^9$&$7.115\times10^{-10}$&$ \le 10^{-3}$&2.05 &
$5.383\times10^{-3}$\\
\noalign{\smallskip}\hline\hline\noalign{\smallskip} 
\end{tabular}
\end{center} 
Note: $\dot m_0$ and $P_0$ vertical mass flow rate and pressure at
the lower boundary $z_0$, $T$ temperature reached at the upper boundary
(sound transition), $\dot M$ rate of inward mass flow in the
corona (=evaporation rate); $\lambda$ fraction of mass carried away by
the wind; quantities $r/r_{\rm S}$ and $\dot M/{\dot
 M_{\rm Edd}}$ scaled to Schwarzschild radius
and Eddington accretion rate. $r, \dot m_0, P_0$ in cgs units.
\end{table*}

\begin{figure}[ht]
\includegraphics[width=7.5cm]{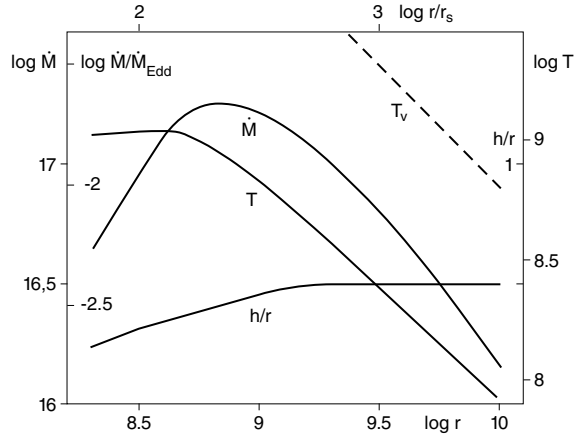}
\caption{
Values of quantities at the inner edge  $r$=$r_{\rm{tr}}$ of the
thin disk for various distances $r$ (in cm). Rate of inward mass flow
$\dot M$(in g/s) in the corona (= evaporation minus
wind loss), maximum temperature in the corona and $h/r$
(h pressure scaleheight), solid lines. Dashed line: virial
temperature. Values of $\dot M$
and $r$ are for a central mass of 6$M_\odot$, scaling with
$\dot M_{\rm{Edd}}$ and $r_{\rm S}$ is indicated.}
\end{figure}

With distance from the black hole the coronal structure changes. 
The main result concerns the evaporation rate, the shape of the
function $\dot M$. 
For small values of $r$ the evaporation rate $\dot M$ increases with
$r$, then reaches a maximum value and decreases further outward.
In Fig. 3 we show the evaporation rate, the temperature reached at
sound transition and the ratio of the pressure scaleheight $h$ to $r$
(maximal temperature and pressure scaleheight from the computed vertical
structure of the corona).
In Table 1 we summarize values of characteristic quantities. Note the large
variation in the values $\dot m_0$ and $P_0$, the values of $P$ and
$\dot m$  at the lower boundary. The evaporation rate is given in
$(M_\odot /y)$ and scaled to the Eddington accretion rate
($\dot M_{\rm{Edd}}=L_{\rm{Edd}}/0.1\rm{c}^2$), the radius in units
of Schwarzschild radius $r/r_{\rm S}$ is added. For different black
hole masses, the evaporation rate can be scaled accordingly (Sect. 6).

To include the evaporation process in the evolution of the accretion disk, e.g.
for black hole X-ray transients, the $\dot M$-$r$ relation for values
$r$ larger than the values in Fig. 3 might be needed.
For such distances much larger than the hump location (Fig. 3,
log$r/r_{\rm S} \ge 3$)
the function $\dot M$ can be approximated as found in the earlier 
investigation by Liu et al. (1995) for accretion in cataclysmic
variables.

In Table 1 we also give the fraction of mass carried away
by the wind. Note that this fraction is of the order 20\% at large
distances $r$ and decreases to very small values inside the maximum of
the function $\dot M$ (Fig. 3). The location of sound transition is
about at
the height 3$r$ for a distance $r=10^{8.8}$cm, the hump
location, and at increasingly larger heights further inward. But their
uppermost layers are unimportant for the energy balance in the corona
because of their very low density.

Disks around black holes can reach inward even to the last stable orbit,
and a new feature appears, a maximum efficiency of evaporation. 
In dwarf nova systems the inner disk edge disk lies much further out
measured in Schwarzschild radii. The maximum of the evaporation
efficiency has important consequences for the inward extension of the
thin disk (see Sect. 5). This maximum is caused by the change in
the physical process that
removes the heat released by friction. For larger radii the
frictional coronal heating is balanced by inward advection and wind
loss. This fixes the coronal
temperature $T$ at about 1/8 of the virial temperature $T_{\rm{v}}$
($\Re T_{\rm v}/\mu=GM/r$).
Downward heat conduction and subsequent radiation 
in the denser lower region play a minor role for the energy loss though 
they always establish the equilibrium density in the corona above the 
disk.  The ratio of the pressure scaleheight $h$ to $r$ is almost
constant for larger $r$ and becomes small closer to the compact
object.

\subsection{The energy balance in the corona}

\begin{figure}[ht]
\includegraphics[width=7.8cm]{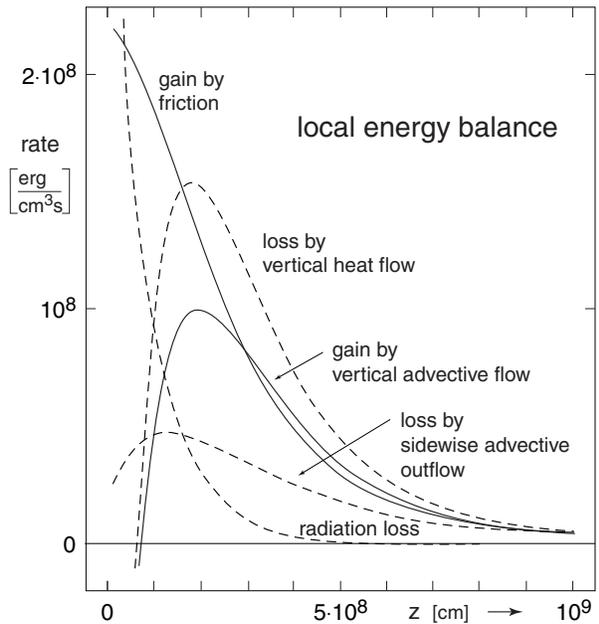}
\caption{Contributions to the local energy balance in the
corona above the thin disk around a 6$M_\odot$ black hole at
log$\,r$=8.8. Solid
lines: gain by friction and vertical advective flow,
dashed lines: loss by radiation, vertical heat flow and sidewise
advective outflow (flow out of the unit volume)}.
\end{figure}

To illustrate the local energy balance we show
in Fig. 4 all contributions separately as functions of height
$z$. There are two terms of energy gain, by friction and by vertical
advective flow. Friction decreases with height with the pressure
(compare the
coronal structure in Fig. 2). The gain by vertical advection results
from the fact that more mass enters from the layer below than leaves
into the layer above. This is due to the net sidewise outflow
out of the unit volume. This overcompensates
the increase of enthalpy with height.
Concerning the losses, the radiation loss is the essential part in
lower layers. It is proportional to the square of density and
decreases steeply towards higher $z$. The loss by vertical heat flow
describes the effect that the heat, released by friction and advection
in higher layers is conducted down to layers where it is radiated
away.

When the temperature has dropped to $10^{6.5}$K at very low height the
downward heat flow is then 10 times larger than the upward advective
flow. The balance is then between conduction and radiation. This leads
to the temperature profile calculated by Shmeleva \& Syrovatskii (1973)
and Liu et al. (1995) and fixes the relation between $F_{\rm c}$ and
$P$ at this temperature used as boundary condition for our calculations.

For log$\,r$=8.8 (Fig. 4) the height-integrated losses by
advective outflow and by
radiation are about equal. This is a particular case. With distance
$r$ the relative size of these two losses varies in a characteristic
way. To show this we have integrated the two losses over the height of
the (flaring) column and normalized by the total release of energy by
friction. Fig. 5 shows the result. The advective loss term includes
the ernergy carried advectively away with the wind. This latter
contribution is also shown separately.

The maximal coronal mass flow rate is reached at that radius where the
character of the solution changes from an advection-dominated flow at
large radii to a radiation-dominated flow at small radii (see Fig. 3).
(This is an intrinsic feature of the solutions as a dimensional analysis of the
global energy balance equation can show).

This behaviour contrasts with that of optically thin accretion flows of
constant mass flow rate $\dot M$. These are advection-dominated
(ADAF) at small radii but become radiation-dominated (Shapiro,
Lightman and Eardley 1976) at large radii (Narayan \& Yi 1995b). The
coronal evaporation, on the other hand, allows for mass exchange
between disk and corona and results in a $\dot M$ varying with $r$
which results in an inversion of that behaviour.

This mass exchange also provides thermal stability to the coronal
evaporation solution. The system responds to a slight increase in
temperature by an increased thermal flux that leads to an increase of
density due to increased evaporation and thereby to a decrease of
temperature by increased radiative cooling.

\begin{figure}[ht]
\includegraphics[width=7.8cm]{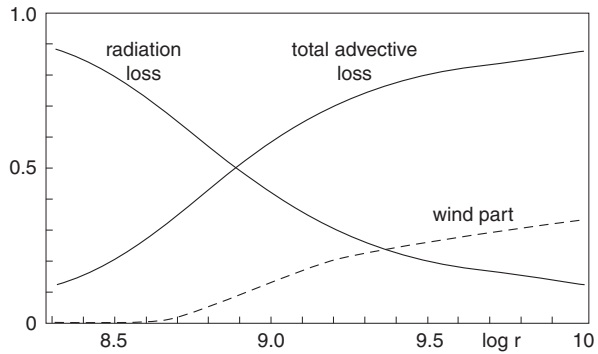}
\caption{Distribution of the total energy released by friction.
Solid lines: fraction going into radiation and fraction
removed by advection (sidewise and via the wind); dashed line: wind
part only. (6$M_\odot$ black hole)}
\end{figure}

\section {The inner edge of the thin disk and spectral transitions} 

If the mass flow rate in the disk
is low, the evaporation rate in the inner disk region may exceed the
mass flow rate. The inner edge of the cool disk then is located where
both rates are equal, that is where all matter brought in by the thin
disk is evaporated. Inside this location 
only a coronal flow exists. This coronal flow provides the supply
for the ADAF towards the compact object. The
higher the mass flow rate in the cool disk the further inward is the
transition to a coronal flow/ADAF. This is the situation in transient
black hole X-ray binaries during quiescence. 
If the mass flow rate exceeds the maximal evaporation
rate, ${\dot M} \ge {\dot M_{\rm{crit}}}$ (the peak coronal mass flow
rate), the inner cool disk cannot be completely evaporated anymore,
instead it reaches inward to the last stable orbit. In high-mass X-ray
binaries fluctuations of the mass flow rate have the same effect.
In Fig. 6 we show how the inward extent of the standard thin disk 
depends on the mass flow rate in the thin disk.
As indicated there the corresponding spectra are hard,
if the hot coronal flow/ADAF dominates, or, soft, if the standard thin
disk radiation dominates near the center. This picture from our modelling 
is the same as the
description of the accretion flow for different spectral states 
presented in the investigation of advection-dominated accretion by
Esin et al. (1997, Fig. 1, except the very high state).

As illustrated
in Fig. 5 the transition between dominant advective losses further out and
dominant radiative losses further in causes the change in the
slope of the evaporation rate at ${\dot M}$= 
${\dot M_{\rm{crit}}}$. Except for the difference
between the sub-virial temperature of the corona and the
closer-to-virial temperature of an ADAF of the
same mass flow rate, a similar critical radius is then predicted by the
``strong ADAF proposal" (Narayan \& Yi 1995b). In general, however, the
strong ADAF proposal results in an ADAF region larger than that which
the evaporation model yields. A detailed discussion
of these spectral transitions was given separately (Meyer et
al. 2000).

\begin{figure}[ht]
\includegraphics[width=8.3cm]{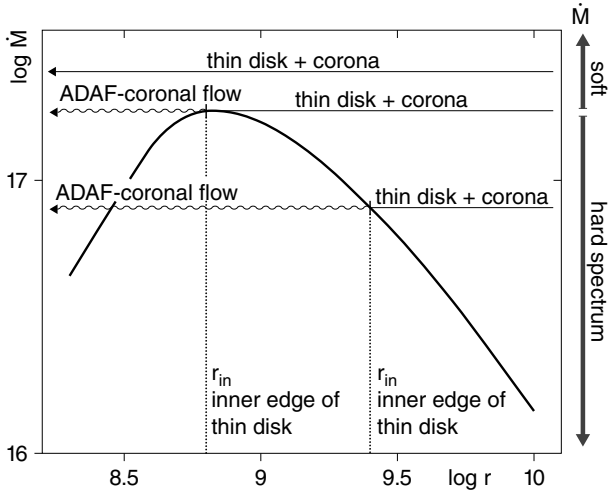}
\caption{
Evaporation rate ${\dot M(r)}$ as in Fig. 3. Inward extension of the standard
thin disk for 3 different mass flow rates ${\dot M}$ in the thin
disk (schematic). Note
that the standard thin disk reaches inward towards the black hole if 
${\dot M} \ge {\dot M_{\rm{crit}}}$. Shown is also the type of spectrum,
soft or hard, related to ${\dot M}$}.
\end{figure}

The main results from our model are the following. The transition
from a thin disk to an ADAF
should occur at about the same luminosity for different systems if their
black hole masses are not too different. The inner edge location
should then either be at a distance of some Schwarzschild radii
(the distance for which $\dot M=\dot M_{\rm{crit}}$) or the thin disk
should reach inward to the
last stable orbit. Several high-mass binaries and black hole
transients show such a systematic behaviour (Tanaka
\& Shibazaki 1996, Tanaka 1999). But in some cases the transition 
seems to be farther inward. In Sect. 9 we compare our results with
observations.

\section{Solutions for different central masses - \\stellar and
supermassive black holes}

Our equations and boundary conditions are invariant against scale
transformations that leave temperature and velocities constant. 
An invariant Kepler velocity implies that $M/r$ is invariant as is
$F_{\rm c}r, Pr, {\rho}r, \dot mr$ and $\dot M/r$.
One thus obtains solutions for another central mass $M_2$ from those
for $M_1$ by multiplying $r,z$ and $\dot M$ values by $M_2/M_1$, and by
dividing $F_{\rm c}, P, \rho$ and $\dot m$ values by the same factor
while keeping velocities and temperatures the same. Since 

\begin {equation}
\dot M_{\rm {Edd}}= \frac{40{\pi}GM}{{\kappa}c},
 \;\;\;r_{\rm S}=\frac{2GM}{c^2}
\end {equation}
are both proportional to mass, the relation between $\dot M$ and $r$
is invariant if plotted in units of Eddington accretion rate and
Schwarzschild radius. In these units Fig. 3 is universal and applies
to both stellar size and supermassive black holes.

The properties of the thin disk underneath do not scale similarly.
This has no effect on the coronal structure, but shows up in the 
spectra belonging to a thin disk + corona for a given mass accretion
rate. 
 
\section {Validity of assumptions}

\subsection{Heat conduction} 
In our modelling classical heat conduction is assumed.
It is only valid if the conductive flux
remains sufficiently small against the maximum transport by free
streaming electrons. The relevant ratio is estimated here as 0.16 at the
radius of maximal mass flow, constant for smaller radii, and decreasing as
$1/{r^{-1/2}}$ for large radii. All estimates are independent of the central
mass. Smaller $\alpha$-values increase the validity of the
approximations.

\subsection{Equilibrium  between electrons and ions}

Our model assumes a one-temperature plasma. Temperature equilibrium between
electrons and ions requires that
the relevant collision timescale between both species remain small 
compared to the heating timescale. Their  ratio is estimated as 0.27 where
the coronal mass flow rate reaches its maximum value, and scales as
$r^{-1/2}$. Thus at smaller radii this approximation brakes down and electrons 
and ions will develop different temperatures.

\section{ Synchrotron radiation, Compton effect}

We have performed estimates on the size of these neglected effects.
Synchrotron losses are proportional to the number density of electrons,
their kinetic energy, and the square of the magnetic field strength. It is 
often assumed that the magnetic field is dynamo produced and that its 
energy density is a fraction $\frac{1}{\beta}$ of the gas pressure.
The ratio of synchrotron losses to thermal plasma radiation then 
becomes a pure function of temperature rising with $T^{3/2}$. The main 
cooling in our solutions occurs in lower layers at about 1/2 of the final 
coronal temperature. Under such circumstances, synchrotron radiation 
remains less than bremsstrahlung for $\beta\geq$ 4. Values for $\beta$
in the literature include $\beta\geq5$ from energy considerations for
ADAFs (Quataert \& Narayan 1999b) and $\beta\leq10$ from direct
numerical MHD calculations (Matsumoto 1999, Hawley 2000).

Compton cooling by photons from the underlying disk depends on the
mass flow rate in the disk. For a rate equal to the maximal coronal
mass flow rate, this Compton cooling approaches about 1/3 of the 
frictional heating. Compton cooling (and heating) of the corona by
radiation from the accretion center depends on the state. In the hard 
state, taking the example of Cyg X-1 (Gilfanov et al. 2000), the
spectrum yields a 
Compton equilibrium temperature of $10^{8.5}$ K. From the flux one
estimates a Compton energy exchange rate comparable with the
frictional heating rate in the corona. This indicates that the true coronal
temperature at the peak of the coronal mass flow rate of Fig. 3 could
be somewhat smaller. A similar result is obtained for Compton
cooling by the central flux during the soft state.  The keV-photons
efficiently cool the corona heated by friction when the temperature
obtained with the pure bremsstrahlung assumption is larger than an
estimated $10^{8.5}$K. 

All these cooling mechanism tend to reduce the
peak temperature and mass flow rate in Fig. 3 and shift the minimal
transition radius somewhat outwards. Our estimates are only rough and
a more quantitative evaluation of these effects has to be done later.

\section{Comparison with observations}

Our model predicts a relation between the evaporation rate and the
inner disk location (function $\dot M$ in Fig. 3 and 6). This fact can
be compared with observations in two ways: [1] for low mass flow rates
in the thin disk a hole should form; the location of the change
from accretion via the thin disk to a coronal flow should be related
to the mass flow rate as predicted; [2] the spectral changes
hard/soft and soft/hard should occur for a predicted mass flow rate
and the location of the inner edge at this time should agree with that
from the model.
To perform test [1] we need to know two quantities, the mass flow
rate in the thin disk and the inner disk edge location. Fitting of the
spectrum based on the ADAF model (for a review see Narayan et
al. 1998) gives values of the mass accretion rate on to the
black hole and therefore also the evaporation rate (= mass flow rate
in the disk at the inner edge). For the location of the inner edge
constraints come from the $\rm{H_{\alpha}}$ emission line. The maximum velocity
indicates the distance of the inner disk edge from the black hole. 
For test [2] the luminosity for which the spectral transition happens
has to be considered, as well as signatures of the inner disk edge.

\subsection{Application to X-ray novae}
For  the X-ray transients A0620-00 and V404 Cyg Narayan et al. (1996,1997a)
derived  mass accretion rates based on spectral fitting of the ADAF model.
These rates were used to perform test [1]. For A0620-00 the
resulting inner edge location agrees reasonably with the value from
the $\rm{H_{\alpha}}$ emission line. For V404 Cyg  a discrepancy was
found (Meyer 1999), but this disappears with the recent results for V404 Cyg 
(Quataert \& Narayan 1999b) and we find agreement also for this system.
The comparison for GRO J1655-40 using the rate from Hameury et al. (1997) and
outer-disk-stability arguments is satisfactory. For details of these
comparisons see Liu et al. (1999).

The agreement of results derived from the disk evaporation on one side
and the ADAF spectral fitting, including different physics for the
description of the innermost region, on the other side, supports both
models.

Further agreement (related to test [1]) between observation and
theoretical
modelling of evaporation comes from the computed disk evolution of
A0620-00 (Meyer-Hofmeister \& Meyer 1999).
The long lasting outburst makes plausible that only little matter 
is left over in the disk. If this is the case 
the evolution during quiescence can be considered independent of the
detailed outburst behaviour. We took into account the evolution of the
thin disk together with the corona. The evaporation determines the
inner edge of the thin disk at each time. The mass flow rate there
is important for the disk evolution and therefore for the outburst
cycle (without careful treatment of the inner edge boundary the
evolution cannot be determined adequately).
The evolution based on the same frictional parameter as taken for
dwarf nova modelling ($\alpha_{\rm{cool}}$=0.05).
leads to the onset of an outburst after the right time and the amount
of matter stored in the disk in agreement with the energy estimated
from the lightcurve. 

Concerning test [2] spectral transitions observed for a few transients
can be considered. The best
observed source is Nova Muscae 1991 (Cui et al. 1997). The transition
occurs around $L_X\approx10^{37}$ erg/s (Ebisawa et al. 1994). These
spectral transitions were modelled based on an ADAF by Esin et al. (1997).  
Our value for the critical mass accretion rate for a 6 $M_\odot$ black
hole, $10^{17.2}$g/s, corresponds to a standard accretion disk
luminosity of about $10^{37.2}$ erg/s, in agreement with
observations.

\subsection{Application to high-mass X-ray binaries}
The accretion rates observed for high-mass X-ray binaries are higher
than those for transients, fluctuations are common. Therefore hard/soft
and soft/hard spectral transitions 
occur much more often than for transients where this phenomenon is
only related to outburst onset or decline.
In our investigation of spectral transitions (Meyer et al. 2000)
we discuss the behaviour of the three persistent (high-mass) black hole
X-ray sources LMC X-1, LMC X-3 and Cyg X-1. The transition
occurs around $L_X\approx10^{37}$ erg/s as also for transients (Tanaka
1999). For the spectral transitions of Cyg X-1 again the ADAF model was
successfully used to describe the observations (Esin et al. 1998).

Our model predicts the inner edge of the thin disk at some 100
Schwarzschild radii for the accretion rate related to the hump in
Fig. 6, that is at the moment of spectral transition. We have
estimated the timescale on which a standard accretion disk at
accretion rates near the critical value $\dot M_{\rm {crit}}$ and at
the related radius can evaporate or form newly and find a value of the
order of days. This agrees with the timescale for the observed spectral
transition of a few days for Cyg X-1 (Zhang et al. 1997).
Frequency resolved spectroscopy of Cyg X-1 (Revnivtsev et al. 1999)
also leads to an inner edge of the thin disk at 150 Schwarzschild
radii. But the existence of a reflecting component indicates
a disk further inward (Gilfanov et al. 1998, Zycki et
al. 1999). For a sample of X-ray transients and high-mass X-ray
binaries Di Matteo \& Psaltis (1999) found, if they relate the rapid variability
properties to those of neutron stars, that the disk reaches down to
about 10 Schwarzschild radii. This would be in contrast to the
prediction from the ADAF based modelling and our evaporation model.
Further work is needed to understand these signatures of a thin disk
so far inward during the hard state.

\subsection{Application to AGN}

Liu et al. (1999) used the results derived in the present paper to 
predict the location of the inner edge of the thin disk around M87 (test [1]).
Using the accretion rate derived by (Reynolds et al. 1996) and an
inner edge of the thin disk from emission lines broadening (Harms et al. 1994) 
they found good agreement between theory and observations.

In connection with new observations further applications of the
evaporation model can be discussed (Liu, Meyer and Meyer-Hofmeister, in
preparation). Recently the
concept of advection-dominated accretion was also applied
to the low-luminosity active galactic nuclei M81 and NGC 4579
(Quataert et al. 1999), NGC 4258 (Gammie et al. 1999) and
low-luminosity elliptical galaxies (Di
Matteo et al. 1999, 2000). It
is not clear why for accretion rates which are different by about less
than a factor of ten (scaled to Eddington accretion) the thin disks
are truncated at $\ge$ $10^4$ Schwarzschild radii or reach inward to 
$\approx$ 100 $r_rm{\rm S}$. Quataert et al. (1999) argued that the
difference might be caused by a different way of mass supply for the ADAF.

\subsection{Spectra from the standard thin disk plus corona}

One might ask whether spectra computed according to our modelling could
be used to test our predictions. Such spectra
would include radiation from an
inner ADAF region and an outer standard thin disk with the transition
at a radius according to our $\dot M - r$ relation. The
spectra determined according to the ADAF model (see e.g. Narayan et
al. 1998) use the observational
constraints for the inner edge of the cool disk and fit the observed
spectra. If the derived rate $\dot M$ and the inner disk location
agree with our $\dot M - r$ relation these already computed spectra are
the spectra that our model will yield.

\section {Discussion of further aspects}

\subsection {Magnetic fields}

The magnetic nature of friction in accretion disks is now generally
accepted, supported by numerical experiments that show the generation
and sustainment of magnetic fields in accretion disks of sufficient
electrical conductivity (for a recent study see Hawley 2000). This
raises the question
whether the neglect of magnetic fields in modelling the corona-disk
interaction may not miss important effects.

Such effects are twofold:
(a) Where gyro-frequency is large compared to collision frequency
electron thermal conductivity is only high along the magnetic field.
This can significantly reduce effective
thermal conductivity if the magnetic field is tangled. This can be
taken into account by reducing the conductivity coefficient $\kappa_0$
by the
average of $cos^2\theta$ over horizontal surfaces, where $\theta$ is
the angle of the magnetic field with respect to the vertical.
(b) Magnetic pressure contributes to the pressure
support. This can be taken into account by increasing the value
$\Re/\mu$ by a factor 1+$\frac{1}{\beta}$ in our equations, where the
$\beta$ is the plasma parameter (which may also include turbulent
pressure).

We have performed an order of magnitude approximation to the
individual terms in the global energy balance. As we will we show in
separate work this allows to model our numerical results and suggests
the following dependence on the parameters $\alpha, \kappa_0$ and
$\beta$.
\[T_0\propto \frac{\alpha^2}{\kappa_0}{\left(1+\frac{1}{\beta}\right)}^3,
\]
\[T_{\rm f} \propto
 \frac{r_{\rm S}}{r}{\left(1+\frac{1}{\beta}\right)}^{-1},
\]
\begin {equation}
\frac{\dot M}{\dot M_{\rm Edd}}\propto \alpha {\kappa_0}^{1/2} 
\left(1+\frac{1}{\beta}\right) {\left(\frac{r}{r_{\rm
 S}}\right)}^{3/2}
 T^{5/2},
\end{equation}
where $T_0$ is the saturation temperature reached at small radii and
$T_{\rm f}$ is the subvirial temperature attained at large radii. The
maximum mass flow rate occurs where the two temperatures $T_{\rm f}$ and
$T_0$ become about equal. This yields the further proportionalities
\[
\left(\frac{\dot M}{\dot M_{\rm {Edd}}}\right)_{\rm {max}}  \propto
\frac{\alpha^3}{\kappa_0^{1/2}}
{\left(1+\frac{1}{\beta}\right)}^{5/2},
\]
\begin {equation}
 \left(\frac{r}{r_{\rm S}}\right)_{\dot M=\dot M_{\rm {max}}} \propto
 \frac{\kappa_0}{\alpha^2}
{\left(1+\frac{1}{\beta}\right)}^{-4}.
\end{equation}
This may serve as a guide on how the results depend on the choice of
these parameter values. For values of $\beta$ discussed in the literature see
Sect. 8.

\subsection {The $\alpha$-dependence}
For our numerical computations we took $\alpha$=0.3. Values discussed
in the literature include 0.1 to 0.3 from spectral fits of ADAF models
to soft X-ray transients (Narayan et al. 1996), $\alpha$=0.25 obtained
from applying detailed calculations to luminous black hole X-ray
binaries (Esin et al. 1997), and $\alpha$=0.1 obtained from global
magnetohydrodynamical simulations of accretion tori (Hawley 2000). 
The above scaling relation shows how our results change if other
$\alpha$-values are chosen.

\subsection{Irradiation}  
X-ray irradiation can cause a corona above the thin disk (e.g. London et
al. 1981). This can interfere with the assumptions underlying the
present model. We note that optically thick scattering layers in
the inner corona can prevent such radiation to reach the disk surface
(e.g. see Schandl \& Meyer 1994). This can become a complex situation
which requires a detailed analysis to clarify under what circumstances
this affects the applicability of our model. In general, models like
ours should be good for black hole systems with ADAF type internal
accretion. Aspects of formation of a disk corona have been
investigated by de Kool \& Wickramasinghe (1999). These calculations are
more detailed with respect to ionization equilibrium than ours, but do
not include thermal conduction, sidewise advection and wind loss in
the energy balance. It is therefore difficult to compare with our results.

\section{Comparison with Honma's model}
Here we have considered the spatial co-existence of disk and corona
and its gradual fading into a corona/ADAF only region at its inner
boundary. Honma (1996) in an apparently very different approach
considered the effect of a turbulent diffusive heat flow outwards from
a hot and mainly non-radiative advection-dominated inner region to an
outercool accretion disk. In his simplified modelling both regions are
treated as one disk with radially varying thickness, and optical
depth. He found a steep transition in the disk thickness and
temperature at a radius that depends on the mass flow rate. In this
radially thin transition layer most of the systems luminosity is radiated.

Though at first sight very different, the two approaches have a basic
physical similarity, though in an inverted geometry: What is vertical
in the disk+corona system is radial in Honma's approach. Heat
generated by friction is conducted outward (downward) from a hot inner
(upper) region to a cool outer (lower) region where it is radiated
away. In both cases this establishes the condition where the cool disk
ends and the hot, ADAF type flow takes over. We note that Honma's
diffusive heat conductivity bears resemblance to the thermal electron
heat conductivity. The rapid increase of the cooling with density
leads to a similarly steep temperature profile in Honma's model as at
the base of the corona in the disk+corona model.

We have calculated the ratio of Honma's effective turbulent heat
conductivity to that of electron thermal heat conductivity and obtain
the value 1/6 for pressure and temperature at the peak of the coronal
mass flow rate in Table 1.

We think that the evaporation model is a more realistic description of
the transition from the cool disk to an corona/ADAF only
flow. In Sect. 2.5 we estimated the radial heat inflow and showed
that it is a minor contribution. Generally, the turbulent outward diffusion
of specific energy reduces the inward advective transport (which is also
diffusive) of energy. Thus the inward flow of energy is somewhat less
than the inward flow of mass times the specific energy. Our
approximate treatment of the inward flows of mass and energy in the derivations
of Eqs. (39) and (46) in effect leads to such a distinction, for
radial power laws as discussed by a factor of 2.

In spite of the very different geometry and simplification Honma's
model captures the same physical effect. His conclusion that his
results shed light on the soft-hard transitions of black hole
accretors is thus substantiated by the disk+corona model, though
quantative differences
exist. In Sect. 5 we have already remarked on where the ``strong ADAF
proposal'' agrees with and where it deviates from the results of the
disk+corona model. This brings the ``strong ADAF proposal'', Honma's
diffusive heat transport model, and the disk+corona model under a common
physical aspect and makes a surprising similarity of some of their
results results understandable. 

\section{Conclusions}
Our model describes the interaction  between the geometrically
thin accretion disk and the corona above.
We have derived simplified equations, which include the relevant
processes determining the accretion via the cool disk together
with the hot coronal flow which yields the matter supply for the ADAF
towards the black hole. We determine the structure of the corona which
is in equilibrium with the cool disk. The equilibrium demands the
evaporation of a certain amount of matter from the cool disk and we
compute the evaporation rate as a function of the distance $r$ from the
central black hole. This gives the rate of accretion of matter
towards the inner region (part of the matter is lost by a wind).
The evaporation efficiency becomes higher with decreasing $r$, but
reaches a maximum at some hundred Schwarzschild radii. For low
accretion rates in the cool disk the evaporation uses up all
matter at the distance where the evaporation rate is equal to the mass
flow rate in the cool disk. The standard thin disk emits a
near black body
spectrum and, for  high enough accretion rates the disk extends
inward to the last stable orbit, the total radiation is soft. But if,
for lower accretion rates
the thin disk is truncated, the coronal flow/ADAF in the inner region
provides a hard spectrum. 
 
These features of disk evaporation apply to galactic black hole X-ray
binaries (X-ray transients, high-mass X-ray binaries) and also to AGN
(truncated disks were found in low-luminosity AGN and elliptical
galaxies). The mass accretion rates derived by spectral fits based on
the ADAF model and the location of the inner disk (deduced from
observations or determined by the fitting)
allow to check the predictions of the model for the location of
transition to the coronal flow. 
These tests show good agreement for most cases of
X-ray transients and the features connected with the maximum evaporation
efficiency (see Sect. 9). We want to point out that our model gives
qualitative results for relations between mass accretion rate,
location of disk truncation and spectral
transitions rather than exact values. The division in an outer
advection-dominated regime and an inner radiation-dominated regime
of the coronal structure gives insight to basic features. 
Computed values for the edge location might differ by
a factor of two to three from the "true'' values. An investigation of
the detailed effect of
simplifications in the equations as well as the effect of the
$\alpha$-value and of magnetic fields will be carried out later.

A truncation of the thin disk farther in than 100 Schwarzschild
radii is in contrast to the predictions of our model.
The reflecting component observed for Cyg X-1 seems to indicate such
an inward extent of the thin disk. But, also for Cyg X-1, another
investigation of observations leads to an inner edge location near
100 Schwarzschild radii (see Sect. 9). This situation needs further
clarification. But also from the theoretical side the
possible existence of an
``interior'' thin disk below a corona in the innermost regions (closer
in than the location of the
evaporation efficiency maximum, compare Fig. 3) needs further investigation.
 
For AGN it was found, based on ADAF spectral
modelling that the truncation of the thin disk can be at very different
distances from the supermassive black hole for mass accretion rates
which differ much less (see Sect. 9). Our model only predicts the
large distances. Also here additional investigations
are needed to understand the physical reason for such divergent
results.

\it{Note added after submission of the manuscript:}
\rm{}In a parallel investigation, very similar to our approach,
R\'o\.za\'nska \& Czerny (astro-ph/0004158, A\&A in press) 
have also studied the coronal structure and the transition to an
ADAF. They take into account the possibility of different temperatures of
electrons and ions and include Compton cooling by the soft radiation from the
underlying disk. In our computed models for the parameters
$r/r_{\rm S}^{3/2} > 25 \dot M_{\rm{Edd}}/\dot M$, that means
log $r/r_{\rm S} \ge 2.25$ (compare Table 1), the
energy equilibration time between electrons and ions is shorter
than the heating time, thus establishing temperature equilibrium
between electrons and ions. For the same parameters also the rate of
Compton cooling by the radiation
from the underlying disk is small compared to the frictional
heating rate and thus negligible.

\end{document}